\documentclass[epjST]{svjour}
\usepackage{amsmath}
\def\n{\text{n}}
\def\s{\text{s}}
\def\jn{\text{jn}}
\def\js{\text{js}}
\renewcommand\Im{\operatorname{Im}}
\begin{document}
\title{Shear viscosity and spin sum rules in strongly interacting Fermi gases}
\author{Tilman Enss}
\institute{Physik Department, Technische Universit\"at M\"unchen,
  85747 Garching, Germany}
\abstract{Fermi gases with short-range interactions are ubiquitous in
  ultracold atomic systems.  In the absence of spin-flipping processes
  the number of atoms in each spin species is conserved separately,
  and we discuss the associated Ward identities.  For contact
  interactions the spin conductivity spectral function
  $\sigma_\s(\omega)$ has universal power-law tails at high frequency.
  We derive the spin $f$-sum rule and show that it is not affected by
  these tails in $d<4$ dimensions.  Likewise the shear viscosity
  spectral function $\eta(\omega)$ has universal tails; in contrast
  they modify the viscosity sum rule in a characteristic way.}%
\maketitle
%

\section{Introduction}

Condensed matter systems near a phase transition generally have
universal low-energy properties, while the high-energy response
depends on non-universal details of the microscopic interaction.
Ultracold atoms provide an important exception: in dilute gases, where
the range of the interaction $r_e$ is much shorter than the mean
particle spacing, also the high-energy properties are universal up to
a cutoff energy $\hbar^2/mr_e^2$ set by the interaction range, which
can be much larger than the Fermi or thermal energies
\cite{bloch2008}.  The correlation functions have characteristic
high-frequency and momentum tails which are controlled by the Tan
contact density $C$
\cite{tan,werner,barth2011,valiente2011,braaten2012}.  This quantity
measures the probability of finding two atoms of different species
near each other.  Together, two atoms can absorb a large kinetic
energy and undergo a boost in opposite directions while conserving
total momentum.  Hence, the high-energy response of the system is
proportional to the density $C$ of such pairs.

In this work we look in particular at the response to a magnetic field
gradient, the spin conductivity $\sigma_\s$, which has recently been
measured \cite{sommer2011a} and provides an example of quantum limited
transport.  Aspects of this are understood within kinetic theory
\cite{bruun2012,enss2012visc}, while a recent strong-coupling
Luttinger-Ward calculation \cite{enss2012spin} explains the spin
diffusion quantitatively and predicts the full frequency dependence of
the spin conductivity $\sigma_\s(\omega)$.  Furthermore, we consider
the response to shear flow, the shear viscosity $\eta(\omega)$
\cite{cao2011,massignan2005,enss2011,romatschke2012} in two and three
spatial dimensions.  The transport coefficients exhibit universal
power-law tails at high frequencies, and we study how these tails
affect the exact sum rules which link the frequency integrated
response to the thermodynamic properties of the system
\cite{enss2012spin,enss2011,pitaevskii2003,taylor2010,hofmann2011,goldberger2012,taylor2012}.
The question of spin transport is connected with the conservation of
the particle numbers $N_\sigma$ of the spin species.  For the case of
a density-density interaction there are no spin-flipping processes and
each $N_\sigma$ is conserved separately.  This implies spin-selective
Ward identities for every spin species, which we will then use to
derive the spin sum rule.


\section{Model and symmetries}
\label{sec:model}

Consider a two-component Fermi gas with contact interaction which
is described by the grand canonical Hamiltonian
\begin{align}
  \label{eq:ham}
  H = \int d\vec x \sum_\sigma \psi_\sigma^\dagger
  \Bigl( - \frac{\nabla^2}{2m} - \mu_\sigma \Bigr)
  \psi_\sigma + g_0 \psi_\uparrow^\dagger \psi_\downarrow^\dagger
  \psi_\downarrow \psi_\uparrow
\end{align}
with mass $m$, chemical potential $\mu_\sigma$ for spin species
$\sigma=\, \uparrow,\downarrow$, and $\hbar\equiv 1$.  At low energies
$s$-wave scattering is allowed only between opposite spins by the
Pauli principle.  The contact interaction $g_0$ leads to ultraviolet
divergences which need to be regularized \cite{bloch2008}.

The interacting Fermi gas \eqref{eq:ham} is invariant under a
$U(1)\times U(1)$ symmetry corresponding to the separate conservation
of $\uparrow$ and $\downarrow$ particle number.  This is readily seen
by coupling a different gauge field to each spin component
\cite{abrikosov1975}.  In the absence of a magnetic field the symmetry
is enlarged to $SU(2)$.  The spin-selective particle number and
current operators can be written as
\begin{align}
  \label{eq:dens}
  j_\sigma^0(x) & = \rho_\sigma(x) = \psi_\sigma^\dagger(x) \psi_\sigma(x) \\
  \label{eq:curr}
  j_\sigma^i(x) & = -\frac{i}{2m} \Bigl( \psi_\sigma^\dagger(x) \partial_i
  \psi_\sigma(x) - \partial_i \psi_\sigma^\dagger(x) \psi_\sigma(x) \Bigr)
\end{align}
where $x=(\vec x,t)$.  These operators satisfy the continuity equation
\begin{align}
  \label{eq:con}
  \partial_t \rho_{\sigma}(x) + \partial_i j_\sigma^i(x)
  = \partial_\mu j_\sigma^\mu(x) = 0
\end{align}
separately for each spin component with conserved particle numbers
$N_\uparrow$ and $N_\downarrow$.  The bare current operator
\eqref{eq:curr} acquires no interaction correction in the case of the
density-density interaction \eqref{eq:ham} since
$[\rho_\sigma(x),\rho_{\sigma'}(y)]=0$ and $[H_\text{int} -
\sum_\sigma \mu_\sigma N_\sigma,\rho_{\sigma'}(x)]=0$
\cite{nishida2007nonrel}.  The continuity equation implies
spin-selective Ward identities which connect the number $J_\sigma^0$
and current $J_\sigma^i$ response functions with the Green's
functions.  These have been derived by Behn \cite{behn1978},
\begin{align}
  \label{eq:ward}
  \partial_\mu \langle Tj_\sigma^\mu(x) \psi_{\sigma'}(y)
    \psi_{\sigma'}^\dagger(y') \rangle
  = \delta_{\sigma \sigma'} \langle T\psi_{\sigma}(y)
    \psi_{\sigma}^\dagger(y') \rangle \left[ \delta(x-y) - \delta(x-y')
  \right] 
\end{align}
or in momentum space
\begin{multline}
  \label{eq:wardmom}
  \omega J_\sigma^0(\vec p,\sigma',\epsilon;
  \vec p+\vec q,\sigma',\epsilon+\omega) 
  - q_i J_\sigma^i(\vec p,\sigma',\epsilon;
  \vec p+\vec q,\sigma',\epsilon+\omega) \\
  = \delta_{\sigma\sigma'} [G_\sigma(\vec p+\vec q,\epsilon+\omega)
  -G_\sigma(\vec p,\epsilon)]
\end{multline}
with Green's functions $G_\sigma^{-1}(\vec p,\omega) = -\omega +
\varepsilon_{\vec p} - \mu_\sigma - \Sigma_\sigma(\vec p,\omega)$ and
the free single-particle dispersion $\varepsilon_{\vec p} = \vec
p^2/2m$.  In particular, there is no response of the $\downarrow$
Green's function to a $q_\mu j_\uparrow^\mu$ perturbation, which is
not immediately obvious from looking at the perturbative
contributions: indeed, Maki-Thompson and Aslamazov-Larkin vertex
corrections \cite{enss2011} cancel exactly in this case.  For $SU(2)$
invariant models there is an additional Ward identity for the
$\sigma^+$ operator \cite{behn1978}.


\section{Spin $\boldsymbol f$-sum rule}
\label{sec:spin}

The correlation functions of number-current $\chi_\jn$ and
spin-current $\chi_\js$ are defined as
\begin{align}
  \label{eq:spincurr}
  \chi_{\jn/\js}(\vec q,\omega) = i\int_0^\infty dt \int d\vec x \,
  e^{i(\omega^+ t-\vec q\cdot \vec x)}
  \langle [j_{\n/\s}^z(\vec x,t),j_{\n/\s}^z(\vec 0,0)] \rangle
\end{align}
in terms of the number and spin current operators $j_{\n/\s}^i(x) =
j_\uparrow^i(x) \pm j_\downarrow^i(x)$ and $\omega^+=\omega+i0^+$.
The corresponding number and spin conductivities in the zero-momentum
limit are defined in terms of the retarded correlation function
\eqref{eq:spincurr} as
\begin{align}
  \label{eq:spincond}
  \sigma_{\n/\s}(\omega) = \frac{\Im \chi_{\jn/\js}(\vec 0,\omega)}{\omega}.
\end{align}
A Kramers-Kronig transformation relates the frequency integral of
$\sigma_{\n/\s}(\omega)$ to the current correlation function at zero
frequency,
\begin{align}
  \label{eq:kk}
  \int_{-\infty}^\infty \frac{d\omega}{\pi} \sigma_{\n/\s}(\omega)
  = \chi_{\jn/\js}(\vec 0,\omega=0),
\end{align}
which is real.  The Kubo formula \eqref{eq:spincurr} can be expressed
in terms of the fermionic Green's and response functions in the
Matsubara formalism as \cite{abrikosov1975,enss2011}
\begin{align}
  \label{eq:spinkubogreen}
  \chi_\js(\vec 0,0) & = -\frac{1}{\beta V} \sum_{\vec p\sigma\sigma'\epsilon_n}
  \frac{\tau_{\sigma\sigma'}^z p_z}{m} \times \tau_{\sigma\sigma'}^z
  J_\sigma^z(\vec p,\sigma',i\epsilon_n) 
  = -\frac{1}{\beta V} \sum_{\vec p\sigma\epsilon_n}
  \frac{p_z}{m} J_\sigma^z(\vec p,\sigma,i\epsilon_n) 
\end{align}
where $\tau_{\sigma\sigma'}^z p_z/m$ is the bare spin-current response
vertex with Pauli matrix $\tau^z$.  This is multiplied with
$J_\sigma^z(\vec p,\sigma',i\epsilon_n)$, the fully dressed current
response function from Eq.~\eqref{eq:wardmom} in the limit
$\omega=0,\vec q\to0$.  The Ward identity \eqref{eq:wardmom} for each
spin component relates the current response function in the static
limit $\omega=0$, $\vec q\to0$ to the Green's function,
\begin{align}
  \label{eq:currward}
  J_\sigma^z(\vec p,\sigma,i\epsilon_n)
  = -\frac{\partial G_\sigma(p,i\epsilon_n)}{\partial p_z}
  = -\frac{p_z}{p} \frac{\partial G_\sigma(p,i\epsilon_n)}{\partial p} .
\end{align}
The Matsubara sum over the Green's function yields the momentum
distribution $-\beta^{-1} \sum_{\epsilon_n} G(p,i\epsilon_n) = n_{\vec
  p\sigma}$ and one obtains
\begin{align}
  \label{eq:16}
  \chi_\js(\vec 0,0)
  & = -\frac 1V \sum_{\vec p\sigma} \frac{p_z^2}{mp} 
  \frac{\partial n_{p\sigma}}{\partial p} 
  = \chi_\jn(\vec 0,0).
\end{align}
The same result is obtained if one considers not the spin-current but
the number-current with bare reponse vertex $\delta_{\sigma\sigma'}
p_z/m$.  The normalized integral over the solid angle $\Omega_d$ yields
$\int d\Omega\, p_z^2 / \Omega_d = p^2/d$ in $d\geq2$ dimensions.
Integration by parts over $p$ then gives
\begin{align}
  \label{eq:partint}
  \chi_{\jn/\js}(\vec 0,0) = -\sum_\sigma \int_0^\Lambda \frac{\Omega_d
    dp\,p^{d-1} }{(2\pi)^d} \frac{p}{md} \frac{\partial n_{p\sigma}}{\partial p}
  = \sum_{\vec p\sigma} \frac{n_{p\sigma}}{m}
  - \frac{1}{md} \frac{\Omega_d}{(2\pi)^d} \sum_\sigma p^d n_{p\sigma}
  \Bigr\rvert_0^\Lambda
\end{align}
where we have explicitly written the ultraviolet momentum cutoff
$\Lambda \sim 1/|r_e|$.  The first term gives the density, while the
second term depends on the cutoff.  For zero-range interactions the
momentum distribution at large momenta is proportional to the Tan
contact density, $n_{p\sigma} = C/p^4$ as $p\to\infty$ in any
dimension \cite{tan,werner,barth2011}.  Hence, the cutoff term
$C\Lambda^{d-4}$ vanishes for $\Lambda\to\infty$ ($r_e\to0$) in any
dimension $d<4$.  In combination with Eq.~\eqref{eq:kk} this completes
the derivation of the particle number and spin sum rule
\begin{align}
  \label{eq:spinsumrule}
  \frac{1}{\pi} \int_{-\infty}^\infty d\omega\, \sigma_{\n/\s}(\omega) 
  = \frac nm
\end{align}
with the total density $n=n_\uparrow+n_\downarrow$.

In the Galileian invariant model \eqref{eq:ham} the number current is
proportional to momentum and cannot decay.  This implies the conservation
of the total number current,
\begin{align}
  \label{eq:momcomm}
  \bigl[H,\int d\vec x\, j_\n^i(x)\bigr] = 0,
\end{align}
and consequently the number conductivity has a sharp Drude peak at
zero frequency,
\begin{align}
  \label{eq:25}
  \sigma_\n(\omega) = \frac nm \pi \delta(\omega).
\end{align}
In contrast, the global spin
current is not conserved because scattering transfers momentum between
$\uparrow$ and $\downarrow$ particles,
\begin{align}
  \label{eq:spincomm}
  [H,\int d\vec x\, j_\s^i(x)] \neq 0,
\end{align}
and the spin conductivity $\sigma_s(\omega)$ has a finite and
nontrivial response at $\omega>0$.  The spin conductivity in 3d has
recently been computed in the Luttinger-Ward formalism
\cite{enss2012spin}: there is a broad Drude peak at low frequencies,
followed by a universal high-frequency tail
\begin{align}
  \label{eq:2}
  \sigma_\s(\omega\to\infty) & = \frac{C}{3\pi(m\omega)^{3/2}} & & (3d)
\end{align}
in accordance with results from the operator product expansion
\cite{hofmann2011}.  Both in two and three dimensions the tail decays
sufficiently fast for the frequency integral \eqref{eq:spinsumrule} to
converge, so again the universal high-energy properties of the zero-range
model do not affect the form of the spin $f$-sum rule in $d<4$ dimensions.


\section{Shear viscosity sum rule}
\label{sec:visc}

The shear viscosity $\eta$ measures the friction of a fluid subject to
a shear flow of both spin species simultaneously (mass flow).  The
real part of the frequency-dependent shear viscosity,
\begin{align}
  \label{eq:eta}
  \eta(\omega) = \frac{\Im \chi_{xyxy}(\omega)}{\omega}
\end{align}
is defined via the retarded stress correlation function 
\begin{align}
  \label{eq:etakubo}
  \chi_{xyxy}(\vec q,\omega)
  = i\int_0^\infty dt \int d\vec x \, e^{i(\omega^+ t-\vec q\cdot \vec x)}
  \langle [\Pi_{xy}(\vec x,t),\Pi_{xy}(\vec 0,0)] \rangle
\end{align}
at zero external momentum, $\vec q=0$.  In general the real shear
viscosity contains an additional contact term proportional to
$\delta(\omega)$ \cite{bradlyn2012}, however in our case of an
interacting Fermi gas at $T>0$ this is canceled by the real part of
$\chi_{xyxy}(\omega=0)$ and does not appear explicitly.  The stress
tensor operator has the off-diagonal components
\cite{enss2011,nishida2007nonrel}
\begin{align}
  \label{eq:Pixy}
  \Pi_{xy} = \sum_{\vec p} \frac{p_xp_y}{m} c_{\vec p-\vec q/2,\sigma}^\dagger
  c_{\vec p+\vec q/2,\sigma}
\end{align}
since the interaction correction vanishes in the zero-range limit
\cite{enss2011,nishida2007nonrel}.  Again a Kramers-Kronig
transformation relates the frequency integral of the viscosity to the
stress correlation function at zero external frequency (static limit),
\begin{align}
  \label{eq:KK}
  \int_{-\infty}^\infty \frac{d\omega}{\pi} \eta(\omega) =
  \chi_{xyxy}(\omega=0).
\end{align}
In analogy with the spin case, the Kubo formula \eqref{eq:etakubo} is
expressed in terms of the stress response function $T_{xy}$ as
\cite{enss2011}
\begin{align}
  \label{eq:etakubogreen}
  \chi_{xyxy}(0) = -\frac{1}{\beta V} \sum_{\vec p\sigma\epsilon_n}
  \frac{p_xp_y}{m} T_{xy}(\vec p,i\epsilon_n) \,. 
\end{align}
In the static limit of external $\omega=0, \vec q\to 0$ the stress
response is determined by the Ward identity associated with momentum
current conservation \cite{polyakov1969},
\begin{align}
  \label{eq:Txyward}
  T_{xy}(\vec p,i\epsilon_n) = -p_x \frac{\partial
    G(p,i\epsilon_n)}{\partial p_y}
   = -\frac{p_xp_y}{p} \frac{\partial G(p,i\epsilon_n)}{\partial p}
\end{align}
and hence
\begin{align}
  \label{eq:5}
  \chi_{xyxy}(0) = -\frac 1V \sum_{\vec p\sigma} \frac{p_x^2 p_y^2}{mp}
  \frac{\partial n_{p\sigma}}{\partial p} .
\end{align}
The normalized integral over the solid angle $\Omega_d$ yields $\int
d\Omega_d\, p_x^2 p_y^2/\Omega_d = p^4/[d(d+2)]$ in $d\geq 2$
dimensions.  Performing an integration by parts as in
Eq.~\eqref{eq:partint} relates the correlation function to the kinetic
energy density $E_\text{kin} = \frac 1V \sum_{\vec p\sigma}
\varepsilon_p n_{p\sigma}$.  The integrals are ultraviolet divergent
and can be regularized by a momentum cutoff $\Lambda$,
\begin{align}
  \label{eq:parts}
  \chi_{xyxy}(0) = \frac 2d\, E_\text{kin} - \frac{1}{md(d+2)}\,
  \frac{\Omega_d}{(2\pi)^d} \sum_\sigma p^{d+2} n_{p\sigma}\Bigr\rvert_0^\Lambda .
\end{align}
Through the momentum distribution $n_{p\sigma} = C/p^4$ for
$p\to\infty$ (see above) the cutoff term depends on the contact
density $C$,
\begin{align}
  \label{eq:chi13}
  \chi_{xyxy}(0) = \frac 2d\, E_\text{kin} - \frac{\Omega_d}{(2\pi)^d}
  \, \frac{C\Lambda^{d-2}}{md(d+2)}.
\end{align}
The kinetic energy density $E_\text{kin}$ can be re-written using the
Tan relations for the internal energy density $\varepsilon$ or the
pressure $P$ as \cite{tan,werner}
\begin{align}
  \label{eq:Ekin}
  E_\text{kin} & = \varepsilon + \frac{C}{4\pi m} \ln
  \frac{\omega_\Lambda}{\varepsilon_B} = P - \frac{C}{4\pi m} +
  \frac{C}{4\pi m} \ln \frac{\omega_\Lambda}{\varepsilon_B} & & (2d)
  \\
  E_\text{kin} & = \varepsilon -
  \frac{C}{4\pi m} \left( \frac 1a - \frac{2\Lambda}{\pi} \right)
  = \frac 32 \left[ P - \frac{C}{4\pi ma} + \frac{C\Lambda}{3\pi^2 m} 
  \right] & & (3d)
\end{align}
with the cutoff energy $\omega_\Lambda = 2\varepsilon_\Lambda =
\Lambda^2/m$ and the two-particle binding energy $\varepsilon_B$.
Then the stress correlation function including the cutoff term in
Eq.~\eqref{eq:chi13} reads
\begin{align}
  \label{eq:10}
  \chi_{xyxy}(0) & = P - \frac{3C}{8\pi m} + \frac{C}{4\pi m} \ln
  \frac{\omega_\Lambda}{\varepsilon_B} & & (2d) \\
  \chi_{xyxy}(0) & = P - \frac{C}{4\pi ma} +
  \frac{4C\sqrt{m\omega_\Lambda}}{15\pi^2m} & & (3d) .
\end{align}
The zero-range interaction leads to universal high-frequency tails
$\eta(\omega) \sim C/(8m\omega)$ in 2d
\cite{hofmann2011,goldberger2012} and $\eta(\omega) \sim
C/(15\pi\sqrt{m\omega})$ in 3d \cite{enss2011,taylor2010}.  These
tails have to be subtracted to make the frequency integral
\eqref{eq:KK} convergent, and one obtains the shear viscosity sum
rules \cite{enss2011,taylor2010,taylor2012}
\begin{align}
  \label{eq:sumrule2d}
  & \frac{2}{\pi} \int_0^\infty d\omega \left[
    \eta(\omega)-\frac{C}{8m\omega} \Theta(\omega-\varepsilon_B)
  \right] 
  = P - \frac{3C}{8\pi m} = \varepsilon - \frac{C}{8\pi m} & & (2d) \\
  \label{eq:sumrule3d}
  & \frac{2}{\pi} \int_0^\infty d\omega \left[
    \eta(\omega)-\frac{C}{15\pi\sqrt{m\omega}} \right] 
  = P - \frac{C}{4\pi ma} = \frac 23 \varepsilon - \frac{C}{6\pi ma} & & (3d).
\end{align}
The universal high-frequency behavior is most clearly seen if one
looks near the quantum critical point at zero density and zero
temperature \cite{nikolic2007,enss2012critical}.  The shear viscosity
in this limit but with the same value of $C$ as in the dense system
has the form \cite{taylor2012}
\begin{align}
  \label{eq:eta0}
  \eta_0(\omega) & =
  \frac{C}{8m\omega} \left(1-\frac{\varepsilon_B}{\omega}\right)^2
  \Theta(\omega-\varepsilon_B)  && (2d).
\end{align}
By subtracting $\eta_0(\omega)$ one arrives a low-energy sum rule
which captures only the finite-density effects \cite{taylor2012}
\begin{align}
  \label{eq:sumrule2dB}
  & \frac{2}{\pi} \int_0^\infty d\omega \left[
    \eta(\omega)-\eta_0(\omega) \right] = P && (2d).
\end{align}

In conclusion, we have argued that zero-range interactions realized in
ultracold atomic systems do not modify the spin $f$-sum rule but lead
to characteristic contact terms in the shear viscosity.


\end{document}